\documentclass[twocolumn]{aastex631}
\usepackage{xfrac,graphicx,color,float,balance}

\newcommand {\cii}{[C{\sc ii}]}
\def\ltsima{$\; \buildrel < \over \sim \;$}
\def\simlt{\lower.5ex\hbox{\ltsima}}
\def\gtsima{$\; \buildrel > \over \sim \;$}
\def\simgt{\lower.5ex\hbox{\gtsima}}

\newcommand{\jwst}{{\it JWST}}

\newcommand {\uJy}{$\mu$Jy}
\newcommand {\um}{$\mu$m}

\newcommand{\msun}{{\rm\,M$_\odot$}}

\newcommand{\lsun}{{\rm\,L$_\odot$}}

\shorttitle{Still no dust in LRDs}
\shortauthors{Casey et al.}

\begin{document}

\title{An upper limit of 10$^6$\msun\ in dust from ALMA observations in 60 Little Red Dots}

\correspondingauthor{Caitlin M. Casey}
\email{cmcasey@ucsb.edu}
\author[0000-0002-0930-6466]{Caitlin M. Casey}
\affiliation{Department of Physics, University of California, Santa Barbara, Santa Barbara, CA 93106, USA}
\affiliation{Cosmic Dawn Center (DAWN), Denmark}

\author[0000-0003-3596-8794]{Hollis B. Akins}\altaffiliation{NSF Graduate Research Fellow}
\affiliation{Department of Astronomy, The University of Texas at Austin, 2515 Speedway Blvd Stop C1400, Austin, TX 78712, USA}

\author[0000-0001-8519-1130]{Steven L. Finkelstein}
\affiliation{Department of Astronomy, The University of Texas at Austin, 2515 Speedway Blvd Stop C1400, Austin, TX 78712, USA}

\author[0000-0002-3560-8599]{Maximilien Franco}
\affiliation{Université Paris-Saclay, Université Paris Cité, CEA, CNRS, AIM, 91191 Gif-sur-Yvette, France}

\author[0000-0001-7201-5066]{Seiji Fujimoto}\altaffiliation{Hubble Fellow}
\affiliation{Department of Astronomy, The University of Texas at Austin, 2515 Speedway Blvd Stop C1400, Austin, TX 78712, USA}

\author[0000-0001-9773-7479]{Daizhong Liu}
\affiliation{Purple Mountain Observatory, Chinese Academy of Sciences, 10 Yuanhua Road, Nanjing 210023, China}

\author[0000-0002-7530-8857]{Arianna S. Long}
\affiliation{Department of Astronomy, The University of Washington, Seattle, WA 98195, USA}

\author[0000-0002-4872-2294]{Georgios Magdis}
\affiliation{Cosmic Dawn Center (DAWN), Denmark} 
\affiliation{DTU-Space, Technical University of Denmark, Elektrovej 327, 2800, Kgs. Lyngby, Denmark}
\affiliation{Niels Bohr Institute, University of Copenhagen, Jagtvej 128, DK-2200, Copenhagen, Denmark}

\author[0000-0003-0415-0121]{Sinclaire M. Manning}\altaffiliation{NASA Hubble Fellow}
\affiliation{Department of Astronomy, University of Massachusetts Amherst, MA 01003, USA}

\author[0000-0002-6149-8178]{Jed McKinney}
\altaffiliation{NASA Hubble Fellow}
\affiliation{Department of Astronomy, The University of Texas at Austin, 2515 Speedway Blvd Stop C1400, Austin, TX 78712, USA}

\author[0000-0002-7087-0701]{Marko Shuntov}
\affiliation{Cosmic Dawn Center (DAWN), Denmark} 
\affiliation{Niels Bohr Institute, University of Copenhagen, Jagtvej 128, DK-2200, Copenhagen, Denmark}

\author[0009-0003-4742-7060]{Takumi S. Tanaka}
\affiliation{Department of Astronomy, Graduate School of Science, The University of Tokyo, 7-3-1 Hongo, Bunkyo-ku, Tokyo, 113-0033, Japan}
\affiliation{Kavli Institute for the Physics and Mathematics of the Universe (WPI), The University of Tokyo Institutes for Advanced Study, The University of Tokyo, Kashiwa, Chiba 277-8583, Japan}
\affiliation{Center for Data-Driven Discovery, Kavli IPMU (WPI), UTIAS, The University of Tokyo, Kashiwa, Chiba 277-8583, Japan}

\begin{abstract}
  By virtue of their red color, the dust in little red
  dots (LRDs) has been thought to be of appreciable influence, whether that dust is
  distributed in a torus around a compact active galactic nucleus
  (AGN) or diffuse in the interstellar medium (ISM) of nascent
  galaxies. In \citet{casey24b} we predicted that, based on the
  compact sizes of LRDs (unresolved in \jwst\ NIRCam imaging),
  detection of an appreciable dust mass would be unlikely.  Here we
  present follow-up ALMA 1.3\,mm continuum observations of a sample of
  60 LRDs drawn from \citet{akins24a}. 
  None of the 60 LRDs are detected in imaging that reaches an average
  depth of $\sigma_{\rm rms}$ = 22\,\uJy.  A stack of the 60 LRDs also
  results in a non-detection, with an inverse-variance weighted flux
  density measurement of $S_{\rm 1.3mm}=$2.1$\pm$2.9\,\uJy. This
  observed limit translates to a 3$\sigma$ upper limit of
  10$^{6}$\,\msun\ in LRDs' dust mass, and
  $\lesssim$10$^{11}$\,\lsun\ in total dust luminosity; both are a
  factor of 10$\times$ deeper than previous submm stack limits for
  LRDs.  These results are consistent with either the interpretation
  that LRDs are reddened due to compact but modest dust reservoirs
  (with $A_{V}\sim2-4$) or, alternatively, that instead of being
  reddened by dust, they have extreme Balmer breaks generated by dense
  gas ($>10^9$\,cm$^{-3}$) enshrouding a central black hole.
\end{abstract}


\section{Introduction}\label{sec:intro}
The discovery of little red dots (LRDs) by JWST has been a perplexing
mystery.  Characterized by compact (unresolved) morphologies and red
rest-frame optical colors (sometimes with a faint, blue component in
the rest-frame UV, giving a `V-shaped' SED;
\citealt{labbe23a,labbe23b,kokorev24a,kocevski24a}), the population of
LRDs at $z>5$ is surprisingly ubiquitous, with volume densities
$\gtrsim$10$^{-5}$\,Mpc$^{-3}$, making up a few percent of the galaxy
population at these epochs.  And yet, a similar population at low-$z$
($z<3$) LRDs -- both compact and red -- seem to be far more rare
\citep{lin25a,bisigello25a,ma25a}.

\begin{figure*}
\includegraphics[width=0.99\textwidth]{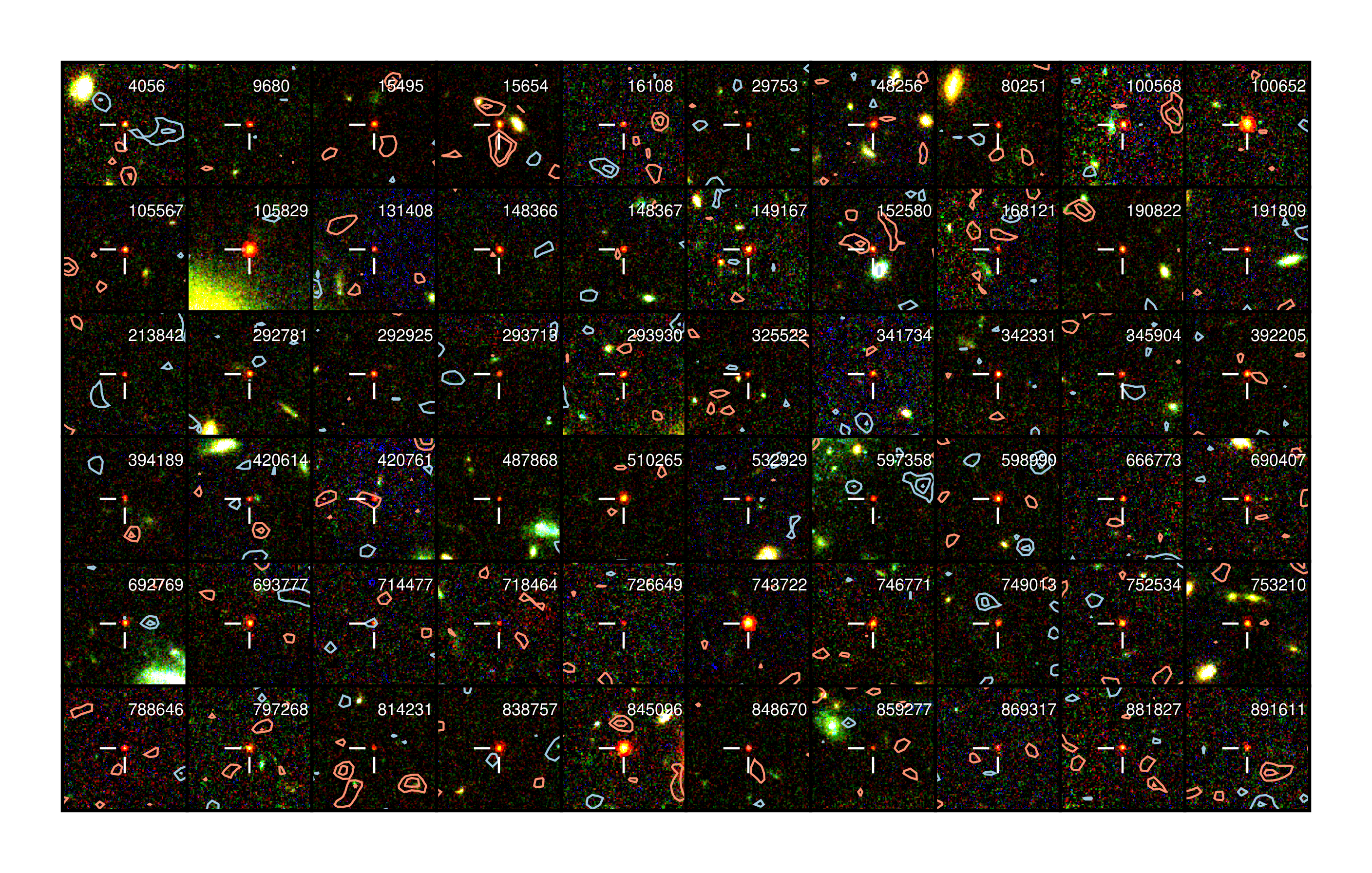}
\caption{JWST 5$''\times$5$''$ tricolor cutouts (F150W, F277W, F444W)
  of our LRD sample overlaid with ALMA dust continuum contours.  Blue
  contours denote positive SNR (integer multiples of $\sigma$
  beginning at 2$\sigma$) and red contours denote negative SNR
  (integer multiples of $\sigma$ descending from -2$\sigma$).  None of
  the 60 LRDs in this program were detected with 1$\sigma$ RMS ranging
  21-23\,\uJy.}\label{fig:gallery}
\end{figure*}

Spectral follow-up efforts have revealed the ubiquity of luminous
active galactic nuclei (AGN) among the LRD population
\citep{matthee24a,greene23a,taylor24a}.  Though the detection of a
Balmer break in some LRDs has led to a hypothesis that LRDs are not
dominated by light from AGN \citep{wang24a}, assuming they are
comprised of highly compact and old stellar populations proves
problematic in a number of ways -- they are excessively dense
\citep{baggen24a} and push the limits of stellar mass assembly at
early times if so \citep{williams23a}.  Alternatively,
\citet{inayoshi24a} posit that LRDs are caused by very dense
($>10^{9}$\,cm$^{-3}$) neutral hydrogen surrounding the accretion disk
of an AGN. This idea is supported by recent observations of LRDs with
extreme Balmer break strengths, exceeding the maximal expected from a
stellar population
\citep{naidu25a,rusakov25a,de-graaff25a,ji25a,taylor25a}.

Whether or not the rest-frame optical emission is dominated by compact
stellar populations or AGN, the red nature of the population has
naturally led to the hypothesis that LRDs are
dust-obscured\footnote{Though we note that the \citet{inayoshi24a}
argument for a black hole enshrouded in dense gas does not explicitly
require {\it any} dust obscuration, and super-Eddington accretion onto
black holes may naturally lead to a redder continuum
\citep{lambrides24a,quadri25a,madau25a}.}.  Significant obscuration
(implied $A_{V}\sim 2-4$) may suggest substantial dust reservoirs
obscuring their light on $\sim$10's or $\sim$100's of parsecs.
\citet{labbe23a} and \citet{akins24a} present some of the first limits
on dust content of LRDs from pre-existing submm datasets in the deep
fields in which their LRDs are found.  No LRDs were detected in those
datasets among hundreds, though their limits were not particularly
sensitive to masses $<$10$^{7}$\,\msun\ (with stacked upper limits of
order $S_{\rm 1.2mm}<$0.1\,mJy at 5$\sigma$).  Recently, deep NOEMA
observations of two $z>7$ LRDs \citep{xiao25a} and separately,
multi-band ALMA observations of two of the brightest LRDs known
\citep{setton25a} result in more stringent non-detections.

In \citet{casey24b} we present the prediction that LRDs' dust masses
are limited to $\sim$10$^{4-5}$\,\msun, with their compact
sizes primarily driving significant attenuation at relatively low
mass.  This was primarily based on earlier submm limits (and in
agreement with the more recent results of \citealt{xiao25a} and
\citealt{setton25a}).  If LRDs were more dust-rich, then they would
have already presented as detections in the submillimeter or at least
show some highly obscured but detectable extended host-galaxy emission
in NIRCam imaging.

In this work we present new ALMA 1.3\,mm continuum follow-up
observations of a large sample of 60 bright LRDs at $z\sim5-8$
selected from \citet{akins24a} to search for anomalously large dust
reservoirs $\gtrsim$10$^{6}$\,\msun\ at $\lesssim$100\,K temperatures.
This constitutes the largest sample of LRDs for which deep dust
continuum measurements have been made.  Section~\ref{sec:data}
presents the LRD sample, ancillary data and ALMA observations,
section~\ref{sec:measurements} presents the measurements, and
section~\ref{sec:summary} discusses the implications of the new
limits.  Throughout we adopt a Planck cosmology
\citep{planck-collaboration20a}, AB magnitudes \citep{oke83a} and a
Chabrier initial mass function \citep[IMF;][]{chabrier03a}.

\section{Sample Selection \&\ Observations}\label{sec:data}

We select LRDs for ALMA observations from the COSMOS-Web
\citep{casey23a} sample of LRDs \citep{akins24a}.  COSMOS-Web covers
more than three times the area of all other JWST deep fields combined
so is particularly sensitive to the rarest, brightest sources.  These
are arguably the most likely LRDs to exhibit a dust detection.  A
third of the field is also covered by MIRI 7.7\um\ imaging, giving
further direct constraints on the rest-frame NIR of the LRD
population.

The LRD selection in \citet{akins24a} can be summarized as (a) a
compactness cut, such that sources are unresolved in F444W, (b) very
red color in NIRCam LW filters, such that $m_{\rm 277}-m_{\rm
  444}>1.5$, and (c) lack of good fit to brown dwarf templates.  It
does not explicitly require a ``V-shaped'' SED in the rest-UV; this
choice was jointly motivated by the lack of understanding of the
physics generating the UV emission and if it is related to the red
rest-optical and depth limitations in COSMOS-Web SW imaging.  While
some may be concerned that the lack of blue rest-frame UV requirement
or sparse filter coverage in COSMOS-Web leaves us more prone to
contamination from, e.g., extreme emission line galaxies, compared to
smaller deep fields with more extensive NIRCam coverage, this concern
is directly addressed by the {\it very} red selection in
[F277W]-[F444W].  The strength of that drop greatly reduces possible
contaminants; the addition of MIRI detection at 7.7\,\um\ further
refines the photometric redshifts of the sample so they have similar
uncertainties as what one might generate using a tight detection of a
Lyman-break in the rest-UV.  Though a small subset, some of the
\citet{akins24a} LRDs have already been spectroscopically confirmed
as broad-line AGN from the COSMOS-3D program \citep{lin25b},
demonstrating a high purity for this LRD photometric selection in
COSMOS-Web.

\begin{figure*}
  \centering
  \includegraphics[width=1.5\columnwidth]{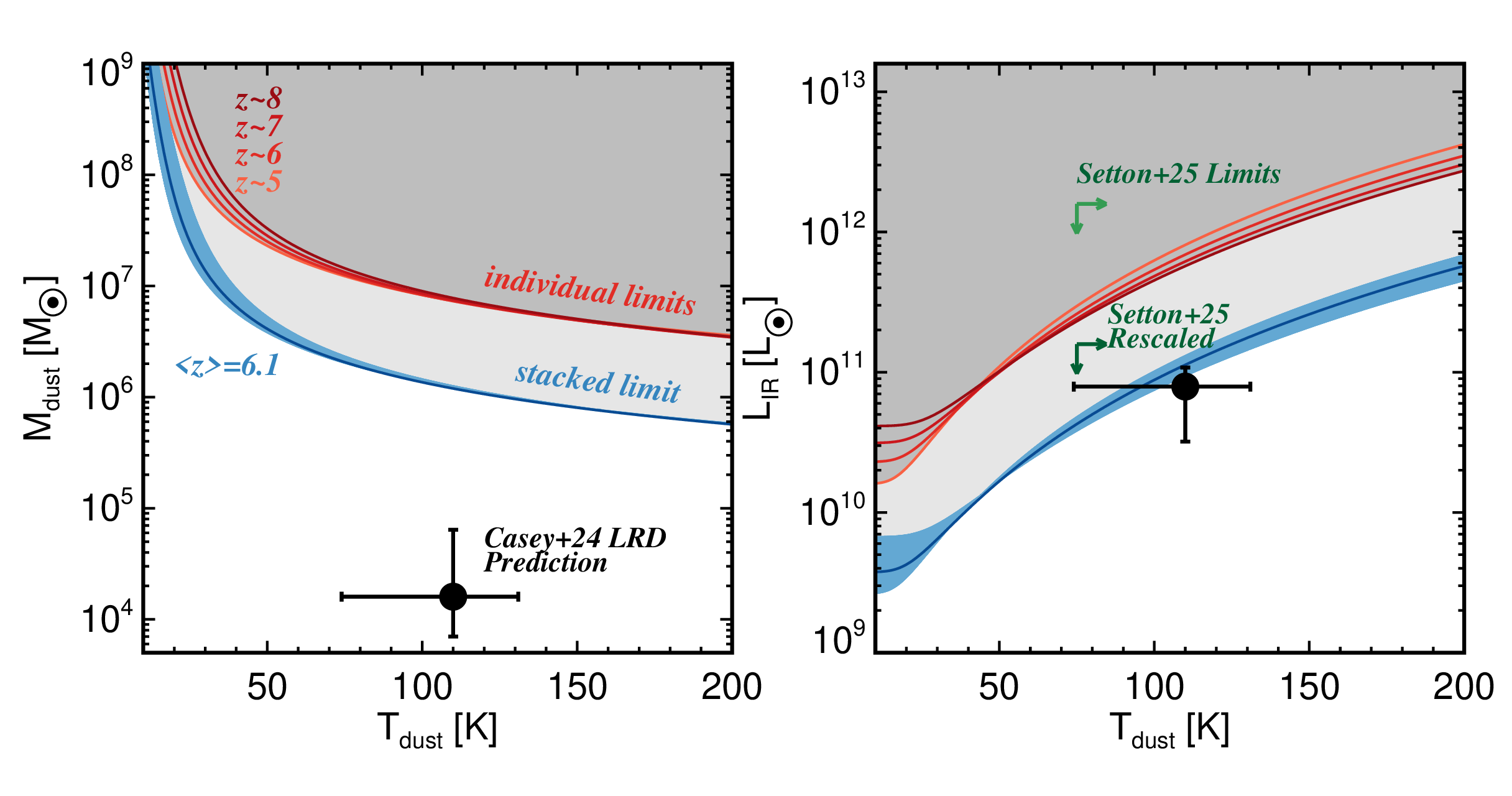}
  \caption{Limits we place on dust mass and IR luminosity based on our
    measurements as a function of dust temperature.  The RMS per
    source ranges from 21--23\,\uJy\ at 1.3\,mm, translating to a
    3$\sigma$ upper limit of 66\,\uJy; this sets an upper limit to the
    dust mass and IR luminosity of LRDs traced out by the red curves.
    We show the redshift dependence of this limit in different shades
    of red, from lightest ($z\sim5$) to darkest ($z\sim8$). The
    limited flux density for the stack, of 10.8\,\uJy\ at 3$\sigma$,
    is traced out by the blue line at the median redshift of the
    sample.  We also show the measured limit to IR luminosity
    ($\lesssim10^{12.2}$\,\lsun) and estimated dust temperature
    ($>$75\,K) from \citet{setton25a}, derived from detailed ALMA and
    JWST/MIRI constraints on two very luminous LRDs (in green); their
    limits rescaled to the typical luminosity of our LRDs is shown in
    dark green.  Despite uncertainty in the underlying dust
    temperature of the dust around LRDs, we place representative upper
    limits from these data of $<10^{7}$\,\msun\ and
    $<10^{12}$\,\lsun\ on individual sources and $<10^{6}$\,\msun\ and
    $<10^{11}$\,\lsun\ on the non-detection from the stack.}
\label{fig:tml}
\end{figure*}

Of the several hundred LRDs found in COSMOS-Web, we select a subset
with the most potential for large dust reservoirs: those that are
reddest (i.e. dustiest, as measured directly via color and through
derived A$_V$ from SED fitting) and the brightest.  Of the 148 LRDs
presented in \citet{akins24a} with MIRI F770W imaging, we apply the
following specific criteria to select a subset for ALMA observations:

$\bullet$ [F444W]$<$27,

$\bullet$ [F770W]$<$26.5 with a detection $>$5$\sigma$, and

$\bullet$ [F277W] - [F444W]$>$1.7.

\noindent
The first criterion selects bright LRDs.  The second (MIRI) criterion
is crucial for anchoring the long-wavelength SED of LRDs, providing
direct evidence that the selected LRDs lack dominant mid-IR powerlaws
(and are thus inconsistent with hot dust tori).  Though these are MIRI
detected, none of the selected sources have SEDs consistent with such
a mid-IR powerlaw; the MIRI detections are also critical for ruling
out brown dwarf contaminants.  The red criterion maps directly to a
high derived absolute magnitudes of attenuation $A_V>2$ from SED
fitting.  Significant attenuation implies that this subset may have
the highest potential to be the dustiest.

The above criteria result in the selection of a sample of 60 LRDs
(about $\sim$40\%\ of the original 148 MIRI-observed LRDs from
\citealt{akins24a}).  Their photometric redshifts span $5<z<8$ (with a
median $\langle z\rangle = 6.1$) and are well-constrained due to their
red [F277W]-[F444W] color and MIRI detections.  This sample size was
motivated by the firm constraints it could set on the results, with a
stacked uncertainty on the resulting flux density $<$5\,\uJy,
translating to a dust mass uncertainty of
2$\times$10$^{6}$\,\msun\ for assumed dust temperature $\sim$100\,K.

ALMA observations were carried out in band 6 (1.3\,mm) as part of
project 2024.1.00708.S (PI: Casey) with 37.3\, hours with the goal of
achieving a 25\,\uJy\,beam$^{-1}$ continuum RMS with $\sim$24\,minutes
on-source.  Observations were split into four tunings to
simultaneously search for serendipitous \cii\ line emission in the
$\sim$1/3 of the sample whose photometric redshifts would place
\cii\ in band 6.  Three of the four tuning setups were executed
covering frequencies 212--220\,GHz, 224--236\,GHz, and 240--244\,GHz
with a representative frequency 228\,GHz ($\lambda=$1.316\,mm).  These
frequencies cover redshift ranges $z=$6.78--6.91, $z=$7.05-7.48, and
$z=$7.63-7.96 for the \cii\ search.  No obvious \cii\ emitters were
found, though we will be able to return to this data once
spectroscopic redshifts are in hand to search for \cii\ using an
informed prior. Spectroscopic redshifts for many will come in the JWST
Cycle 3 program COSMOS-3D (\#\ 5893, e.g. \citealt{lin25b}) and 
the Cycle 4 program EMBER (\#\ 7076, PI: Akins).

Data were obtained spanning 11-Dec-2024 through 14-Jan-2025 and were
reduced using the CASA pipeline version 6.6.1.17.  The continuum
sensitivity reaches per-source depths spanning $\sigma_{\rm rms}$ =
21.4--23.1\,\uJy\,beam$^{-1}$ with mean noise of
22.3\,\uJy\,beam$^{-1}$.  The synthesized beam averages to
0$\farcs$81$\times$0$\farcs$65 using natural weighting.

\begin{figure*}
\includegraphics[width=0.99\textwidth]{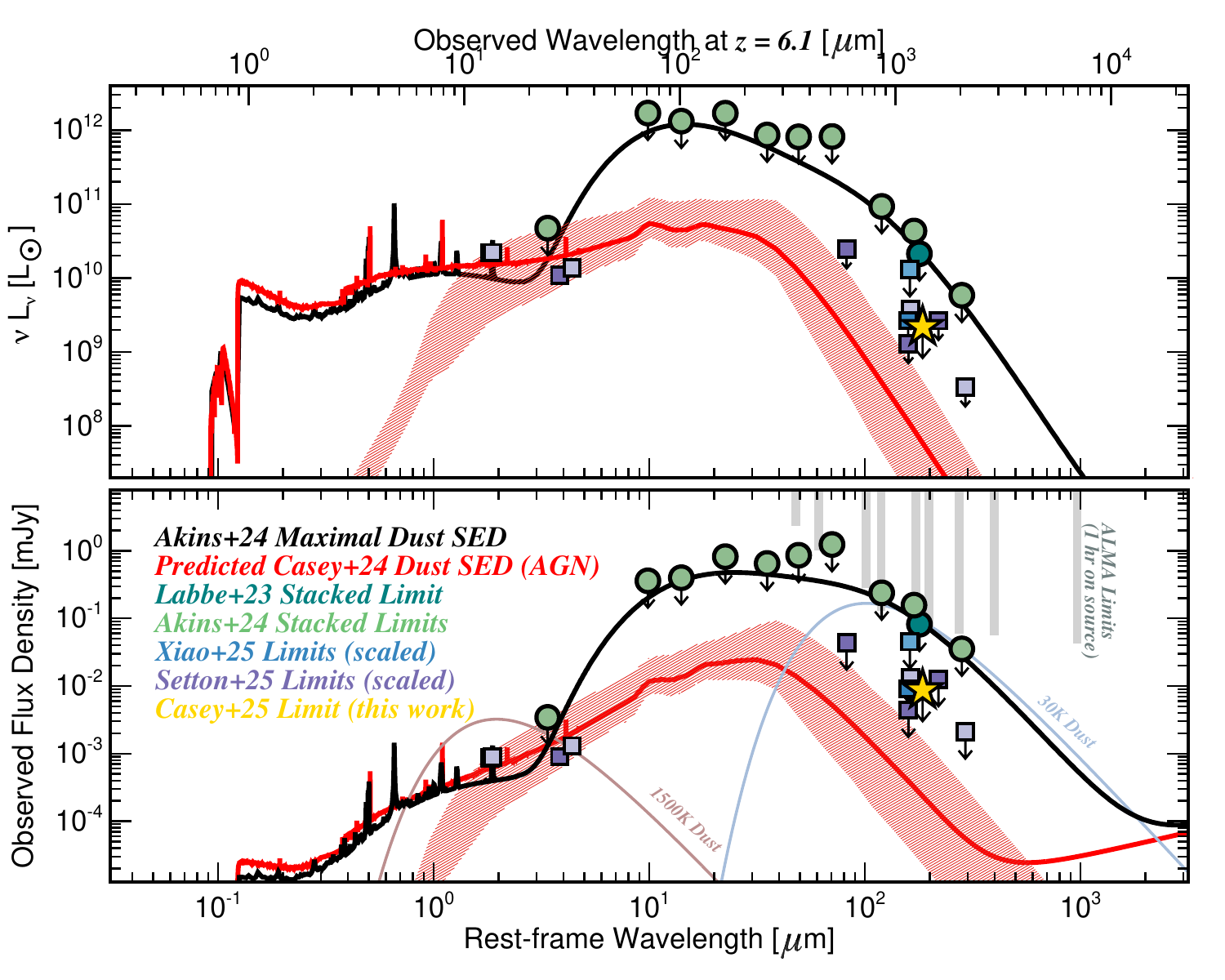}
\caption{Our measured constraints on the stack of 60 LRDs (gold star)
  compared to other literature limits of dust emission around LRDs.
  The top panel plots the SED constraints in $\nu L_\nu$ and the
  bottom panel in observed flux density. Other limits from the
  literature are shown in dark teal \citep{labbe23a}, light green
  \citep{akins24a}, blue \citep{xiao25a}, and purple
  \citep{setton25a}.  The \citeauthor{xiao25a} and
  \citeauthor{setton25a} limits are set for two different LRDs each:
  light blue represents ID9094 at $z=7.0388$, dark blue is ID2756 at
  $z=7.1883$, light purple is RUBIES-BLAGN-1 at $z=3.1034$ and dark
  purple for A2744-45924 at $z=4.4655$. Because these limits are all
  for unusually bright LRDs, we have scaled their limits down by a
  factor of 3, 10, 10, and 10 (respectively) to draw a more direct
  comparison to the LRDs targeted in our program.  Two optical through
  radio SEDs are shown: the maximal SED from \citet{akins24a}, showing
  the maximum allowable dust SED around LRDs given the limits
  available (black), and the predicted LRD dust SED from
  \citet{casey24b} in red.  On the bottom panel we overplot simple
  modified black body SEDs with temperatures of 30\,K and 1500\,K for
  comparison.  Our new limit for this aggregate sample is fully in
  line with the predicted low luminosity dust emission.}
\label{fig:sed}
\end{figure*}

\section{Derived Limits on Dust Mass}\label{sec:measurements}

None of the 60 LRDs were detected in dust continuum.  With typical RMS
of 22\,\uJy, this gives a typical per-source 3$\sigma$ (5$\sigma$)
upper limit of $S_{\rm 1.3\,mm}<$67\,\uJy\ ($<$111\,\uJy).  The
measured SNR of each source's central unresolved beam ranges from --2.0
to 1.8, fully consistent with expectation from the null hypothesis,
i.e. that none have discernible millimeter emission.  The gallery of
LRDs covered by this program are shown in Figure~\ref{fig:gallery},
with ALMA contours overlaying JWST tri-color images.

Beyond the non-detection of individual sources, we also stack our data
in the image plane.  This also results in a non-detection.  The
measured constraint on the flux density from the stack using
inverse-variance weighting is $S_{\rm
  1.3\,mm}=2.1\pm2.9$\,\uJy\ (0.72\,$\sigma$).  We note this is not
appreciably different from a direct, unweighted stack of $S_{\rm
  1.3\,mm}=1.9\pm3.2$\,\uJy\ (0.58$\sigma$) because the RMS varies by
less than 10\%\ across the full sample.

Thanks to the negative K-correction, both dust mass and IR luminosity
constraints derived from a single flux density measurement are
largely insensitive to redshift.
We calculate 3$\sigma$ upper limits on dust masses in our sample by
implementing Equation~1 of \citet{casey19a}, which is 
dust mass corrected for CMB heating.
Because the LRDs span appreciably high redshifts $5\lesssim
z\lesssim8$ the calculation of dust mass (or luminosity) requires
accounting for the effect of CMB heating where the CMB temperature is
non-negligible in the first 1\,Gyr of cosmic time.  CMB heating
matters proportionally more for dust temperatures close to the CMB
temperature, $T<50$\,K.  Eq~1 of \citet{casey19a} does require a
number of adoptions: the dust mass absorption coefficient \citep[for
  which we use the measurement at 450\um\ rest-frame of
  1.3$\pm$0.2\,cm$^2$\,g$^{-1}$;][]{li01a}, the dust emissivity index
$\beta=2$, and an assumption of the dominant temperature of the dust
mass in LRDs.  The temperature is the dominant source of uncertainty.

Figure~\ref{fig:tml} shows the temperature dependence of the dust mass
limit and IR luminosity limit derived from our constraints on the
1.3\,mm flux density of LRDs.  Gray regions of parameter space are
ruled out by our observations.  Red curves show the typical 3$\sigma$
upper limit per source where we use the representative $\sigma_{\rm
  rms}=22\,$\uJy.  Blue curves show the stacked limit of
10.8\,\uJy\ at 3$\sigma$ for the full sample.  Dust mass limits are
largely insensitive to temperature above $\sim$50\,K. \citet{casey24b}
present a discussion of why cold dust temperatures ($\lesssim$50\,K)
are not expected in LRDs given their compact morphologies.  Note that
an independent estimate of dust-heating by super-Eddington accreting
AGN predicts temperatures $\sim$140\,K \citep{mckinney21a}.  Given the
lack of strong temperature evolution on our derived limits above these
temperatures, we quote the upper limit on dust mass of
$<10^{7}$\,\msun\ on individual sources and $<10^{6}$\,\msun\ on the
stack for the full sample.

We can similarly place limits on IR luminosity, though these are more
temperature dependent than dust mass, given that our single-band
1.3\,mm measurement likely probes the Rayleigh-Jeans tail of emission
for temperatures $\sim$100\,K. Our observations probe
140--220\,\um\ in the rest-frame.  Figure~\ref{fig:tml} shows this
temperature dependence.  If we use the dust temperature of
$\sim$114\,K estimated in \citet{casey24b}, which is consistent with
other independent theoretical estimates on the dust temperature
\citep{li24a}, then we can roughly place upper limits on the IR
luminosity of LRDs at $<10^{12}$\,\lsun\ (at 3$\sigma$) for individual
measurements and at $<$10$^{11}$\,\lsun\ for the stack.

\section{Summary}\label{sec:summary}

We present ALMA band 6 (1.3\,mm) continuum observations of a sample of
60 little red dots drawn from \citet{akins24a}.  None of the LRDs have
dust continuum detections; similarly, a stack of all ALMA data also
results in a non-detection.  The RMS of these observations,
$\sim$22\,\uJy, places a dust mass limit for individual LRDs at
$<$10$^{7}$\,\msun\ at 3$\sigma$ and a limit of $<$10$^{6}$\,\msun\ at
3$\sigma$ from the stack.  These limits are consistent with our
previous work \citep{casey24b} which argues that the dust mass
reservoir around LRDs should be of order $\sim10^{4-5}$\,\msun\ based
on their compact sizes and relatively modest attenuation.  If we adopt
the predicted dust temperature of LRDs of $\sim$100\,K from
\citet{casey24b}, which is consistent with the temperature
independently estimated by \citet{li24a}, we place a limit on the IR
luminosity of LRDs at $\sim$10$^{11}$\,\lsun\ (3$\sigma$, for the
stack).  We show the stacked limit in context with the average (and
predicted) SED for LRDs in Figure~\ref{fig:sed}.

Our results are consistent with two other literature works that placed
limits on dust in LRDs in the literature: \citet{xiao25a} present
NOEMA 1.3\,mm constraints for two bright $z>7$ LRDs, and
\citet{setton25a} present multi-band ALMA follow-up of two
exceptionally luminous LRDs at $z=3-5$.  Both our measurements and
those in \citet{xiao25a} and \citet{setton25a} place the most
stringent constraints on long wavelength dust emission for the
population, roughly an order of magnitude deeper than previous
millimeter continuum constraints from \citet{labbe23a} and
\citet{akins24a}.  Our constraints are largely complimentary to the
other stringent limits by constraining a large sample of 60 LRDs as
opposed to very deep observations of 2 (particularly bright) LRDs
each.

To-date, no LRD has been firmly identified through its millimeter
emission in continuum \citep[only one has a tentative detection of
  neutral gas in the millimeter; A2744-45924 at $z=4.46$ presented
  in][ whose continuum limits come from
  \citeauthor{setton25a}]{akins25b}. If predictions hold, detection
via dust continuum will likely not happen without significant
sensitivity upgrades to ALMA's receivers.  Future far-infrared
missions such as PRIMA will be able to reach deeper stacking limits at
$\lambda_{\rm obs}\sim30-200$\,\um, and may be capable of directly
detecting warm dust emission in the brightest LRDs like A2744-45924.

It could be, however, that the predicted dust content might not even
be there if LRDs' reddening originates from physics beyond
dust-obscuration entirely.  In the emerging picture that LRDs may be
dominated by a population of low-mass black holes cocooned in a shroud
of dense, ionized gas accreting at super-Eddington luminosities
\citep[e.g.][]{inayoshi24a,naidu25a,rusakov25a,de-graaff25a}, then the
'reddening' would instead be caused by at atmosphere of dense gas
relatively devoid of dust.  If that hypothesis holds, it logically
follows that their submillimeter emission would be even less
significant than the already meager predictions.

\begin{acknowledgements}
CMC thanks the National Science Foundation for support through grants
AST-2009577 and AST-2307006 and to NASA through grant JWST-GO-01727
awarded by the Space Telescope Science Institute, which is operated by
the Association of Universities for Research in Astronomy, Inc., under
NASA contract NAS\,5-26555.  HBA acknowledges support from the
Harrington Graduate Fellowship at UT Austin, and HBA and ORC thank the
National Science Foundation for support from Graduate Research
Fellowship Program awards.  This project has received funding from the
European Union’s Horizon 2020 research and innovation programme under
the Marie Skłodowska-Curie grant agreement No 101148925.
\end{acknowledgements}

\end{document}